\documentclass[twocolumn, a4paper]{revtex4}
\usepackage{graphicx}
\usepackage{dcolumn}
\usepackage{amsmath}
\begin{document}

\hsize\textwidth\columnwidth\hsize\csname@twocolumnfalse\endcsname

\title{Controlled single electron transfer between Si:P dots}

\author{T. M. Buehler, V. Chan, A. J. Ferguson, A. S. Dzurak, \\F.
E. Hudson, D. J. Reilly\footnote{Now at Now at Dept. Physics,
Harvard University, Cambridge MA 02138, USA}, A. R. Hamilton and R.
G. Clark} \affiliation{Centre for Quantum Computer Technology
Schools of Physics and Electrical Engineering, University of New
South Wales, NSW 2052, Australia }
\author{D. N. Jamieson, C. Yang, C. I. Pakes and S. Prawer}
\affiliation{Centre for Quantum Computer Technology School of
Physics, University of Melbourne, VIC 3010, Australia }

\begin{abstract}
\vspace{0.5cm} \noindent \small{We demonstrate electrical control
of Si:P double dots in which the potential is defined by nanoscale
phosphorus doped regions. Each dot contains approximately $600$
phosphorus atoms and has a diameter close to $30$ nm. On
application of a differential bias across the dots, electron
transfer is observed, using single electron transistors in both
dc- and rf-mode as charge detectors. With the possibility to scale
the dots down to few and even single atoms these results open the
way to a new class of precision-doped quantum dots in silicon.}
\end{abstract}

\maketitle

Considerable progress has recently been made towards spin-based
quantum computing in semiconductors, most notably with experiments
that probe and control electron spin coherence in GaAs quantum dots
\cite{ElzermanNature,PettaScience,KoppensScience}. An outstanding
challenge for semiconductors remains coherent control of single
electron spins bound to individual $\mathrm{^{31}P}$ donors in
isotropically pure silicon $\mathrm{^{28}Si}$ which promises to
allow extremely long coherence times. In Kane's original
scheme~\cite{Kane_nature} the qubits were defined by nuclear spin
states of $\mathrm{^{31}P}$ dopants in Si. Since then, other Si:P
schemes have been proposed based on both
spin~\cite{vrijen,Frieson_PRB,Sousa_PRA,Hill_PRA} and
charge~\cite{Hollenberg_PRB}.

To assess the feasibility of Si:P qubits we have constructed double
dot structures in silicon using phosphorus ion implantation with
surface control gates and rf-single electron transistor (rf-SET)
readout circuitry (Fig. 1a). These many-electron dots have a
metallic density of states separated by a barrier, enabling periodic
sequential tunneling between dots upon application of a differential
bias to the surface gates. The SETs provide non-invasive detection
of this charge motion in contrast to direct transport measurements
from previous work \cite{Fay_MNE}. The ability to control and
detect, on rapid timescales, the motion of a single electron between
these Si:P buried atom dots addresses a number of formidable
challenges associated with Si:P qubits.


To construct the Si:P dots we use electron-beam lithography (EBL)
to pattern 30nm apertures in a polymethyl-methacrylate (PMMA)
ion-stopping resist, laterally defining the doped regions. A 14
keV $\mathrm{^{31}P}$ ion beam implants the 600 dopants per hole
through a 5nm $\mathrm{SiO_{2}}$ barrier layer to a mean depth of
20nm into the substrate~\cite{Jamieson_APL}. Damage created during
implantation is repaired via a rapid thermal anneal to
1000$^{\circ}$C for 5 seconds, sufficient to activate the P
donors~\cite{McCamey_semi} but limiting their diffusion to
$\sim$1nm based on bulk rates~\cite{Fahey_RMB}. The control gates
are deposited using EBL, followed by the two
$\mathrm{Al/Al_{2}O_{3}}$ SETs, fabricated using a bilayer resist
and double-angle metallization process.

To control the transfer of electrons in the double dot structures,
the symmetry of the double well potential is adjusted by applying a
differential voltage to gates $\mathrm{S_{L}}$ and $\mathrm{S_{R}}$.
The resulting charge distribution in the double dot is then
determined by monitoring the source-drain current (or reflected rf
amplitude in the case of rf operation) $\mathrm{I_{SD}}$ of either
of the two SET charge detectors on the device.

Fig. 1b plots the SET current as $\mathrm{V_{SL}}$ is swept over a
100mV range for a device with a dot separation d=100nm. Throughout
these measurements, compensation voltages on additional gates keep
the SETs at operating points of maximum sensitivity. We show four
periods of a sawtooth waveform, characteristic of single electron
transfer between the two dots (inset Fig. 1a). The sawtooth can be
understood as a steady polarization of the system until it becomes
energetically favourable for an electron to be transferred. The
transfer events occur with an average period of
$\mathrm{V_{SL}}$=24mV over a wide range of gate voltages. We note
that these events are not perfectly periodic and additional charge
noise is also present, discussed below.

In Fig. 1c, we plot ($\mathrm{dI_{SD}/dV_{SL}}$) as a function of
the two gate voltages $\mathrm{V_{SL}}$ and $\mathrm{V_{SR}}$. The
locus of the charge transfer events is highlighted by the dark
lines, indicating high transconductance. These follow linear
trajectories and show a period of 24mV in $\mathrm{V_{SL}}$ and
224mV in $\mathrm{V_{SR}}$, indicating a stronger capacitive
coupling of the double dot to the left symmetry gate
$\mathrm{S_{L}}$ than to $\mathrm{S_{R}}$. This can be explained
by lithographic alignment errors between the surface gates and the
buried double dot. We have measured five double dot devices, with
dot separations of d=80nm, 100nm and 150nm. All show the
characteristic quasi-periodic sawtooth seen in Fig. 1b, but in
each case the relative coupling capacitance and period varies due
to differing misalignment. We have calculated the capacitance
matrix for our devices for varying displacement between the
surface gates and buried double dots, finding values consistent
with our data~\cite{Lee_Nanotech}. Displacements of order 50nm can
modify the capacitance or transfer period by up to an order of
magnitude. Improvements in lithographic alignment procedures could
address this issue in future devices.

Whilst, in principle, a single SET is sufficient to detect charge
transfer within the double dot (or Si:P qubit), by correlating the
output from two SETs~\cite{Buehler_APL} it is possible to reject
spurious events resulting from charge motion in the substrate, or
within the SETs themselves. Figs. 2a,b show simultaneous measurement
of both SETs on the DQD with d=100nm, taken after thermal cycling to
room temperature. A simple correlation achieved by multiplying the
two transconductance signals for the right and left SETs is shown in
Fig. 2c for $\mathrm{V_{RC}}$ = -41mV. The correlation signal
exhibits a sharp peak whenever a charge transfer event occurs,
whilst uncorrelated events are suppressed. Correlated detection now
provides strong evidence that the charge transfer is occurring
\textit{in the region between} the two SET couplers.

The magnitude of the charge signal $\Delta$q induced on a SET island
due to the electron transfer is a key parameter characterizing the
time required for the readout process. We find $\Delta$q = 0.035e
for a double dot with d=100nm (Fig. 2d) and $\Delta$q = 0.038e for a
d=150nm sample, consistent with values obtained from a simplified
model of our structure.

\begin{figure}
\begin{center}
\includegraphics[width=7cm]{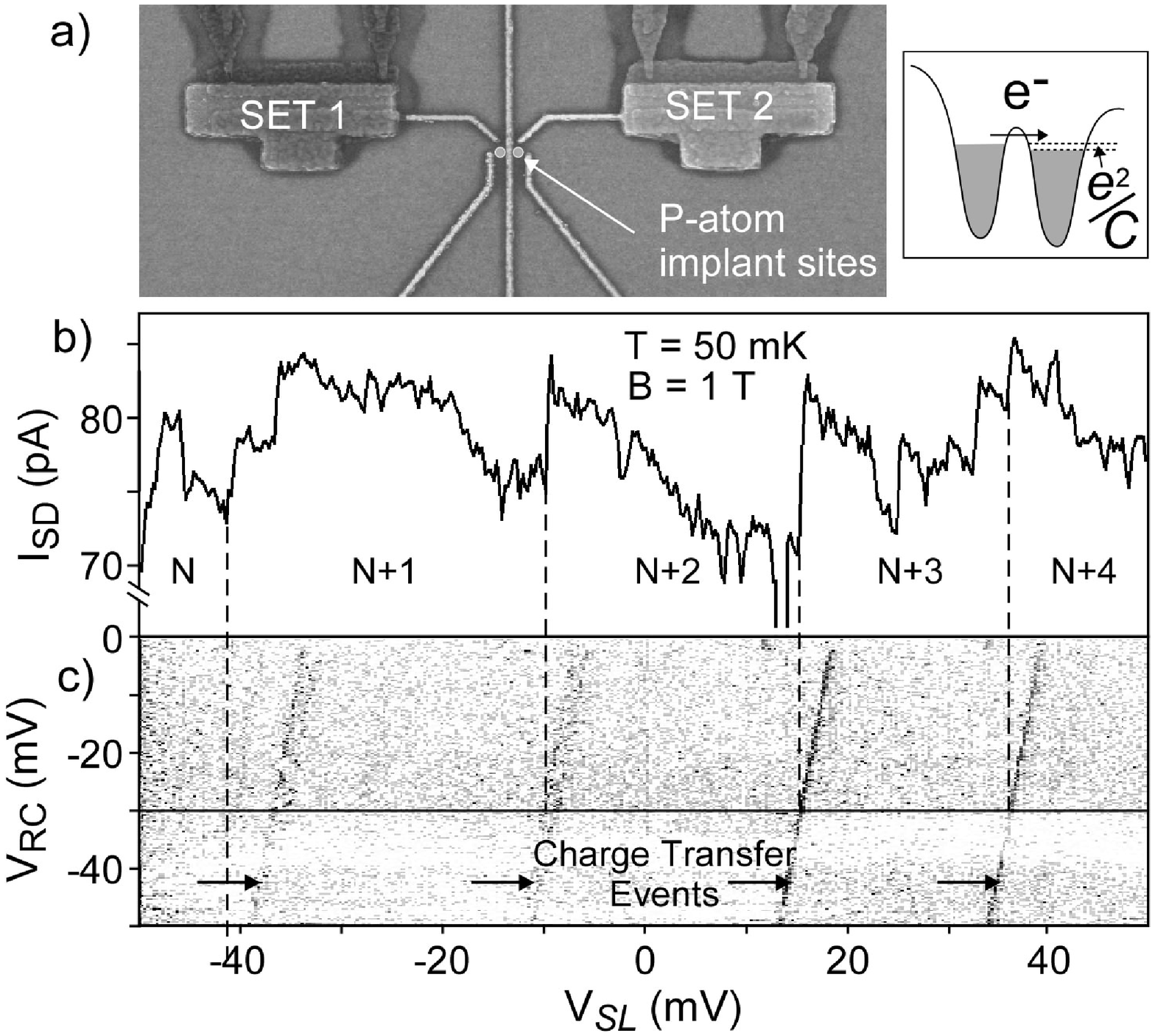}
\caption{\small \textbf{a)} SEM image of a Si:P double-dot device
with three control gates and two SETs. White circles indicate
implant regions. \textbf{b)} SET current $\mathrm{I_{SD}}$ as a
function of $\mathrm{V_{SL}}$ with $\mathrm{V_{SR}}$ = -30mV
(horizontal line in \textbf{c}) for a double dot device with d =
100nm. \textbf{c)} SET transconductance $\mathrm{dI_{SD}/dV_{SL}}$
(intensity plot) as a function of $\mathrm{V_{SL}}$ and
$\mathrm{V_{SR}}$ for the same sample. B = 1T was applied to
suppress superconductivity at T = 50 mK.}\vspace{-0.9cm}
\end{center}
\end{figure}

\begin{figure}[t]
\begin{center}
\includegraphics[width=7cm]{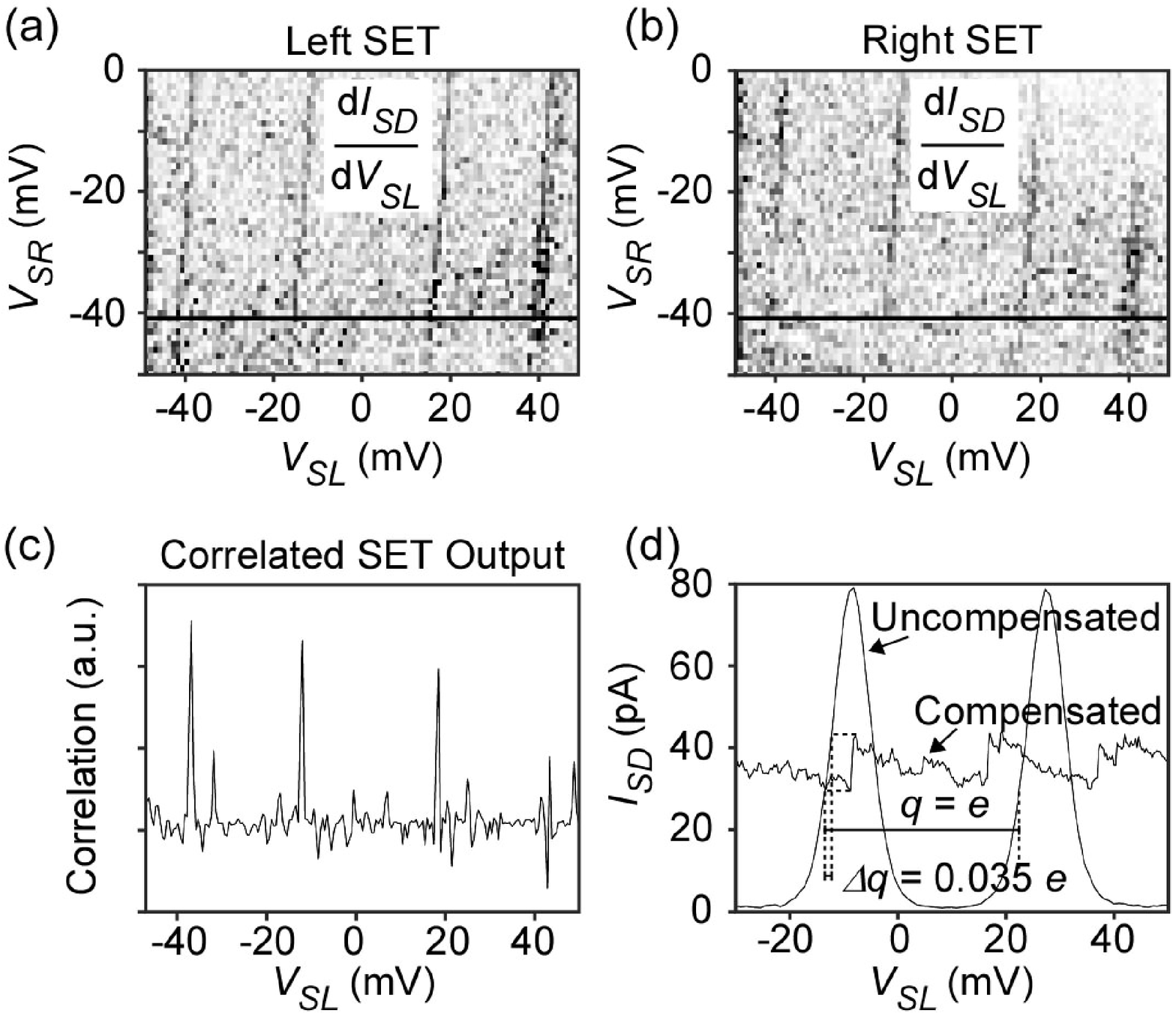}
\caption{\small Electron transfer events observed in
$\mathrm{dI_{SD}/dV_{SL}}$ of \textbf{a)} left SET, and
\textbf{b)} right SET, as a function of $\mathrm{V_{SL}}$ and
$\mathrm{V_{SR}}$ for a double dot sample with d=100nm.
\textbf{c)} Correlated SET output plotted against
$\mathrm{V_{SL}}$ (horizontal line in a,b). Sharp peaks correspond
to single electron transfer events. \textbf{d)}~Determination of
induced charge  $\Delta$q due to a single electron transfer event.
For the inter-dot transfer event we deduce  $\Delta$q = 0.035e. T
= 50 mK, B = 1T.}\vspace{-0.9cm}
\end{center}
\end{figure}

Readout of qubits requires detectors that operate on time scales
faster than the qubit relaxation (or excitation) time. Towards this
goal, the radio-frequency SET (rf-SET) offers near quantum-limited
sensitivity~\cite{Schoelkopf_science, Korotkov_PRB} with switchable
back-action~\cite{Johansson_PRL}. Large (100 MHz) measurement
bandwidths can be achieved by embedding the SET detector in a LC
matching network, mapping dc conductance to reflected rf power. In
rf measurements with B = 0T, the SET source-drain bias data (Fig.
3a) exhibit a typical Josephson quasiparticle spectrum. To maximize
the detection signal we bias the SET to a region of small
differential resistance ($V_{SD} \simeq 4\Delta$), where the
sensitivity exceeds $\mathrm{10^{-5}e/\sqrt{Hz}}$. Fig. 3b shows the
time-domain response of an rf-SET to a voltage step applied to a SET
bias gate which induces a charge of $\Delta$q = 0.1e on the SET
island. Fig. 3c shows the output signal from one rf-SET on a double
dot device with d = 150nm as a function of time while a differential
bias is applied between the control gates $\mathrm{V_{B}}$ and
$\mathrm{V_{SR}}$. Here again, the rf signal shows a characteristic
sawtooth response, indicating periodic transfer of single electrons
between the buried phosphorus dots. The signal to noise ratio of the
rf response is greater than the dc data in Fig. 1a since there is
reduced 1/f charge noise due to the shorter timescale.

A number of control experiments were also carried out on the double
dots and related devices. The integrity of the ion-stopping mask was
confirmed by fabricating nanocircuitry as in Fig. 1a but omitting
the apertures for substrate doping. These devices showed no evidence
of periodic charge motion in the substrate. The metallic density of
states in the dots was confirmed by the observation of ~75 electron
transfers in the voltage range $\mathrm{V_{SL}}$ = [-900mV, 900mV],
all with a period close to 24 mV and all with the same slope in gate
bias (Fig. 1a, 2a,b). Small variations in periodicity, seen in all
data, are believed to be related to internal physical and electronic
structure of the dots. Control devices with no implants, or with
silicon implants, but with the same SET and gate structures were
measured under identical conditions. No periodic transfer was seen
in these although random charge transfer features, most likely due
to electron traps in the substrate, were observed~\cite{Furlan_PRB}.

\begin{figure}[t]
\begin{center}
\includegraphics[width=7cm]{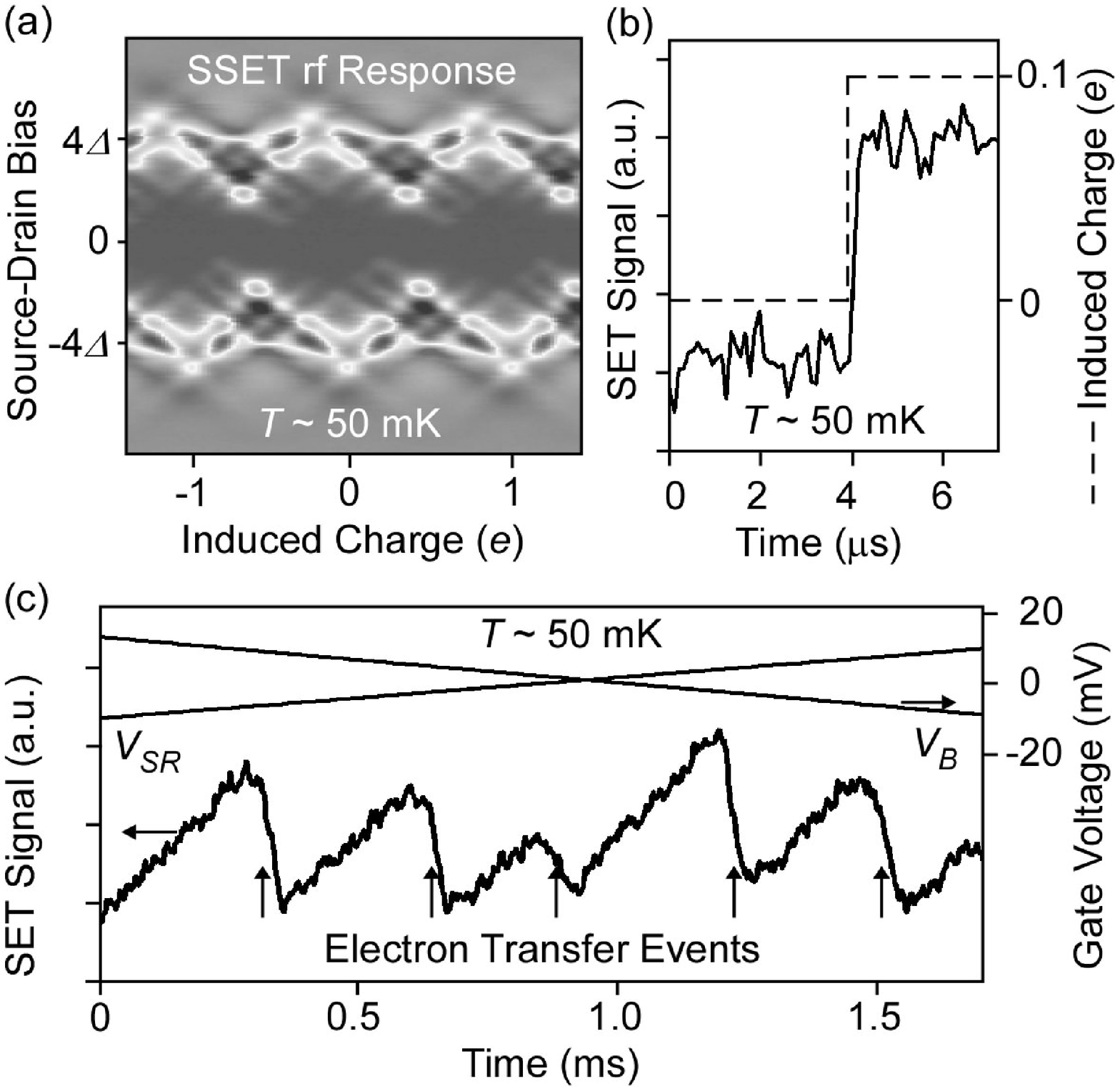}
\caption{\small Measurements using rf-SET detection. \textbf{a)}
Bias spectroscopy of a rf-SET in the superconducting state (B =
0T) where $\mathrm{\Delta}$ is the superconducting gap for Al.
\textbf{b)} Single-shot response of the rf-SET to a small step in
gate voltage creating an induced charge of 0.1e at the SET island.
\textbf{c)}Sawtooth signal from periodic electron transfer in a
double dot with d = 150nm, observed in the rf-SET signal (left
axis scale) as a function of time while a differential bias
voltage (right axis scale) is applied to control gates B and
$\mathrm{S_{R}}$. Data is an average of 16 traces, each with
acquisition time 1.7 ms. }\vspace{-0.9cm}
\end{center}
\end{figure}

The results here demonstrate a gate-controlled Si:P double dot
system with the facility for fast measurement of inter-dot electron
transport. To our knowledge these Si:P double dots are the only type
defined by localized doping of silicon. They can also be reduced to
single atom dots using single-ion implantation~\cite{Jamieson_APL}.

The Si:P double dot devices demonstrated here represent a critical
step towards silicon-based quantum computing, being configured with
control and readout circuitry at the scale required for a two-atom
Si:P charge qubit. The fast detection of single electron transfer
over a distance of order 100nm provides good prospects for qubit
readout, while correlated twin-SET detection offers additional
immunity from materials-related charge noise. Future experiments
will involve microwave spectroscopy on both DQD and two-P-atom
devices to map out the energy levels and determine $T_{1}$ and
$T_{2}^{*}$.

This work was supported in part by the Australian Research
Council, the Australian Government, the U.S. National Security
Agency, the Advanced Research and Development Agency and the U.S.
Army Research Office under contract no. DAAD19-01-1-0653. We
acknowledge E. Gauja, R. Starrett and A. Greentree for useful
discussions.

\end{document}